\documentclass[pre, aps, superscriptaddress, showpacs, twocolumn]{revtex4}
\usepackage{bm, epsfig}
\begin{document}
\title{Exploring classical phase space structures of nearly integrable
and mixed quantum systems via parametric variation}
\author{Nicholas R. Cerruti}
\affiliation{Department of Physics, Washington State University, Pullman,
Washington 99164-2814}
\author{Srihari Keshavamurthy}
\affiliation{Department of Physics, Washington State University, Pullman,
Washington 99164-2814}
\affiliation{Department of Chemistry, Indian Institute of Technology,
Kanpur U. P. 208 016, India}
\author{Steven Tomsovic}
\affiliation{Department of Physics, Washington State University, Pullman,
Washington 99164-2814}
\date{\today}

\begin{abstract}
The correlation between overlap intensities and level velocities has been
introduced as a sensitive measure capable of revealing phase space
localization.  Previously applied to chaotic quantum systems, here we
extend the theory to near-integrable and mixed quantum systems.  This
measure is useful in the latter cases because it has the ability to
highlight certain phase space structures depending upon the perturbation
used to parametrically vary the Hamiltonian.  A detailed semiclassical
theory is presented relating the correlation coefficient to the phase
space weighted derivatives of the classical  action.  In the process, we
confront the question of whether the Hannay-Ozorio de Almeida sum
rules are simply extendable to mixed phase space systems.  In addition,
the $\hbar$-scalings of the correlation coefficient and relevant
quantities are derived for nearly integrable systems.  Excellent
agreement is found between the theory and the results for integrable
billiards as well as for the standard map.

\end{abstract}

\pacs{05.45.Mt, 03.65.Sq}
\maketitle

\section{Introduction}

System response to parametric variations or perturbations is of great
importance.  It is a powerful experimental practice from which new
information about a system can be extracted that is not generally
available by other means, especially for complex systems,
i.e.~disordered, interacting many-body, and/or simple chaotic systems.
External parameters such as electromagnetic fields, temperatures, applied
stresses, etc., are controllable and often suitable for this purpose.
Some specific examples include conductance fluctuations in quantum
dots~\cite{quantumdots}, and variations in eigenmodes due to shape
deformations of microwave cavities~\cite{sridhar:94} and vibrating
plates~\cite{schaadt:99}.  Intramolecular vibarational energy
redistribution~\cite{gruebele:00} in molecules is another example
wherein the technique of parametric variations can lead to useful insights
regarding the key perturbations responsible for strong mode mixings.  Molecular
spectroscopy in external fields is an area of current interest and the
response of a molecular system to external fields can be usefully analyzed
from the parametric perspectives~\cite{fraser:94}.

In the extreme chaotic or disordered limit, there exist universalities
in system response to perturbations~\cite{simon:93}, and the eigenstates
respect the ergodic hypothesis~\cite{berry:77}, i.e.~no phase space
localization; there is only a scale to extract.  On the other hand, the
response of integrable or near-integrable systems is far richer being
system dependent, and the Husimi distributions of the eigenstates are
confined to classical tori~\cite{takahashi:86,kurchan:89}.  A
perturbation is likely to give rise to subsets of eigenlevels/states that
respond very similarly as a group, and the parametric quantities thereby
carry far more information.  For mixed systems both regular and chaotic
trajectories form phase space structures and the most obvious, naive
initial starting point would be to attempt to analyze each structure
separately.  At this level of approximation, using the appropriate
semiclassical theory for regular and chaotic regions involves
the respective integrable and chaotic Hannay-Ozorio de Almeida sum rule
expressions~\cite{hannay:84} weighted by the relative fractions of phase
space volume.  We find instead that the intermittent behavior of the
unstable orbits prevents this simple picture from applying.

A measure was introduced by one of us (S.T.)~\cite{tomsovic:96} that
reveals phase space localization of quantum eigenstates.  Originally
introduced for chaotic systems, the measure correlates the changes in the
eigenenergies due to a perturbation (termed ``level velocities'') with
the overlap intensities between the eigenstates and a probe state, which
is often usefully chosen to be a coherent state.  More recently, it has
been identified as contributing to the scale associated with the fidelity
in the weak perturbative regime~\cite{ct2}.  For integrable systems phase
space localization is the norm, not the exception.  There and in the
near-integrable regime, the localization is associated with tori,
resonance zones and stable periodic orbits.  The correlation measure has
been shown to highlight different features of classical phase space
depending upon the perturbation~\cite{keshavamurthy:02}.   Several years
ago Weissman and Jortner~\cite{weissman:82} performed a similar study
involving the Husimi distributions and parametric changes in eigenenergy
levels.

This paper is organized as follows:  we begin by developing the
semiclassical theory of the correlations for near-integrable systems
where we also remark on the more general theory for the level velocities
in mixed systems.  The correlations have been previously studied
in highly excited rovibrational states in molecules where multiresonant
Hamiltonians are applicable~\cite{keshavamurthy:02}.
Section~\ref{semiclassical} gives the semiclassical theories of
the level velocities, strength functions and overlap intensity-level
velocity correlation coefficients.  The semiclassical theory is compared
with various results from integrable billiards and the standard map which
has the entire range of dynamics from integrable to fully chaotic.  We
finish with a discussion of the similarities and distinctions between
near-integrable and chaotic systems and some comments about mixed systems.

\section{Preliminaries}

Consider a near-integrable quantum system governed by a smoothly
parameter-dependent Hamiltonian $\hat{H}(\lambda)$ with classical
analog $H({\bf I}, {\bm \theta}; \lambda)$ where
\begin{equation}
H({\bf I}, {\bm \theta}; \lambda)
= H_0({\bf I}) + \lambda H_1({\bf I}, {\bm \theta})
\end{equation}
Without loss of generality, the phase space volume of the energy surface,
$V$, is taken to be a constant as a function of $\lambda$.  This ensures
that the eigenvalues do not collectively drift in some direction in
energy, but rather wander locally.  The parameterized strength function is
given by
\begin{eqnarray}
      S_\alpha(E, \lambda) &=& {1 \over 2 \pi \hbar} \int^\infty_{-\infty}
        e^{iEt/\hbar} \langle \alpha | e^{-i\hat{H}(\lambda)t/\hbar}
        | \alpha \rangle dt \nonumber \\
      &=& \sum_n p_{\alpha n}(\lambda) \delta(E - E_n(\lambda))
\end{eqnarray}
where $p_{\alpha n}(\lambda) = |\langle \alpha | E_n(\lambda) \rangle |^2$.
$S_\alpha(E, \lambda)$ is the Fourier transform of the autocorrelation
function of a normalized initial state $| \alpha \rangle$.  Ahead
$\overline{S}_\alpha(E, \lambda)$ will denote the smooth part resulting
from the Fourier transform of just the extremely rapid initial decay due
to the shortest time scale of the dynamics (zero-length trajectories).
We will take $| \alpha \rangle$ to be a Gaussian wave packet because of
its ability to probe `quantum phase space', but other choices may be
useful depending on the circumstances.  In one spectroscopic application,
we found it to be advantageous to take $| \alpha \rangle$ as a momentum
state~\cite{keshavamurthy:02}.

The overlap intensity-level velocity correlation coefficient is
defined as
\begin{equation}
      {\cal C}_\alpha(\lambda) \equiv {\left \langle p_{\alpha n}
      {\partial E_n(\lambda) \over \partial \lambda} \right \rangle_E
      \over \sigma_\alpha \sigma_E}
\end{equation}
where $\sigma^2_\alpha$ and $\sigma^2_E$ are the local variances of
intensities and level velocities, respectively.  The brackets denote
a local energy average in the neighborhood of $E$.  It weights most the
level velocities whose associated eigenstates possibly share common
localization characteristics and measures the tendency of these levels
to move in a common direction.  In this expression, the phase space
volume remains constant so that the level velocities are zero centered
(otherwise the mean must be subtracted), and ${\cal C}_\alpha(\lambda)$
is rescaled to a unitless quantity with unit variance making it a true
correlation coefficient.  The set of states included in the local energy
averaging can be left flexible except for a few constraints.  Only
energies where $\overline{S}_\alpha(E, \lambda)$ is roughly a constant
can be used or some intensity unfolding must be applied.  The energy
range must be small so that the classical dynamics are essentially the
same throughout the range, but it must also be broad enough to include
enough eigenstates for statistical purposes.

\section{Semiclassical Dynamics}
\label{semiclassical}

We develop a theory based upon semiclassical dynamics for the overlap
correlation coefficient of regular states by examining its individual
components, the level velocities and intensities.  The theory for
the variance of the level velocities involves the trace formula for
near-integrable systems developed by Ullmo, Grinberg and
Tomsovic~\cite{ullmo:96}.  The near-integrable trace formula is used
rather than the simpler integrable trace formula of Berry and
Tabor~\cite{berrytabor:77} since it correctly describes the
contributions of short, unstable periodic orbits that occur after a system
breaks integrability.  The level velocities are derivatives and thus far
more sensitive to the perturbation than the intensities.  Hence, it is
more important to accurately describe the level velocities.   The
intensity variances for near-integrable systems are given by the
integrable result as long as the coherent state is not near a resonance
zone.

It turns out that a general result for the level velocity
variance in mixed systems cannot be obtained simply by applying the
appropriate Hannay-Ozorio de Almeida sum rule for the effectively
regular and chaotic trajectories separately.  The ``effective'' trajectories
are not easily defined since they do not merely correspond to the
trajectories within the KAM islands and the chaotic sea,
respectively.  Although, it is observed that the near-integrable result
for the level velocities is in good agreement with the calculated results
before the break up of the last KAM torus.

\subsection{Level velocities}
\label{velocities}

By adapting a method employed by Berry and Keating~\cite{berry:94} for
classical chaotic systems with the topology of a ring threaded by quantum
flux and later generalized for all continuous time chaotic
systems~\cite{leboeuf:99,cerruti:01}, the variance of the level
velocities can be obtained.  The details for near-integrable
systems are in appendix~\ref{lv} where the variance of the level
velocities is found to be
\begin{eqnarray}
      \sigma^2_E
        &\approx& {4 \pi \epsilon \over V \hbar}
        \sum_{\bf M} {}^{\prime} \left \langle A_{\bf M}^2
        \left\{ \left( \partial \overline{S}_{\bf M}
        \over \partial \lambda \right)^2
        [J_0^2(s_\lambda) + \tilde{a}^2 J_1^2(s_\lambda)] \right. \right.
        \nonumber \\
      &+&  \left. \left( \partial \Delta S_{\bf M}
        \over \partial \lambda \right)^2
        \left[ {\tilde{a}^2 \over 4} (J_0(s_\lambda) - J_2(s_\lambda))^2
        + J_1^2(s_\lambda) \right] \right\} \nonumber \\
      &\times& \left.
        \exp \left( {-2 \epsilon T_{\bf M} \over \hbar}\right) \right \rangle_E
\end{eqnarray}
where the $\epsilon \rightarrow 0$ limit is understood, and the amplitude
factor is determined by the integrable system,
$H_0({\bf I})$
\begin{equation}
      A_{\bf M} = {1 \over 2 \pi |{\bf M}|^{(d + 1) / 2} \sqrt{|K_{\bf M}|}}
\end{equation}
The prime on the sum excludes the ${\bf M} = 0$ term.
${\bf M}$ labels the rational tori and is a $d$-dimensional vector
with positive integer components whose classical
paths are those which at time $t$ have returned to the same point on their
torus after making $M_1$ circuits of coordinate $\theta_1$, $M_2$ circuits
of coordinate $\theta_2$, etc.~\cite{berrytabor:77}. $K_{\bf M}$ is the
scalar curvature matrix of the energy contour and $T_{\bf M}$ is the
period of the unperturbed orbit on the resonant torus.  $J_0(z)$ and $J_1(z)$
are the standard Bessel functions.  By the
Poincar\'{e}-Birkhoff theorem only two periodic orbits
survive the break-up of a rational torus.  One orbit is stable and the other
is unstable with actions $S_s$ and $S_u$ and stability matrices $M_s$ and
$M_u$, respectively.  Hence, we define
\begin{equation}
      \overline{S}_{\bf M}(\lambda) \equiv {S_u + S_s \over 2}
      \ \ \ \mathrm{and} \ \ \ \Delta S_{\bf M}(\lambda)
      \equiv {S_u - S_s \over 2}
\end{equation}
with $s_\lambda = \Delta S_{\bf M}(\lambda) / \hbar$ and
\begin{equation}
      \tilde{a}(\lambda) \equiv {\kappa - 1 \over \kappa + 1}
\end{equation}
where
\begin{equation}
      \kappa = \left( -{\det (M_u - {\bf 1})
      \over \det (M_s - {\bf 1})} \right)^{1/2}
\end{equation}
In the case $\kappa \rightarrow 1$ then $\tilde{a}(\lambda) \rightarrow 0$
and the spectral staircase reduces to Ozorio de Almeida's
result~\cite{ozorio:86}.

The summation over rational tori can be accomplished by applying the
Hannay-Ozorio de Almeida sum rule for integrable systems~\cite{hannay:84}
\begin{equation}
      \label{eq:sum_rule}
      \sum_{\bf M} {}^{\prime} A_{\bf M}^2 \cdots
      \rightarrow {V \over 2 \pi} \int {dT \over T^2} \cdots
\end{equation}
Thus,
\begin{eqnarray}
      && \sigma^2_E
      \approx {2 \epsilon \over \hbar} \int {1 \over T^2}
      \left \langle  \left\{ \left( \partial \overline{S}_{\bf M}
        \over \partial \lambda \right)^2
        [J_0^2(s_\lambda) + \tilde{a}^2 J_1^2(s_\lambda)] \right. \right.
        \nonumber \\
      &&  + \left. \left. \left( \partial \Delta S_{\bf M}
        \over \partial \lambda \right)^2
        \left[ {\tilde{a}^2 \over 4} (J_0(s_\lambda) - J_2(s_\lambda))^2
        + J_1^2(s_\lambda) \right] \right\}  \right \rangle_{\bf M}
        \nonumber \\
      && \times \exp \left( {-2 \epsilon T \over \hbar}\right) dT
      \end{eqnarray}
For rational tori, the change of the classical action is proportional to
the period of motion.  We can think of this as a ``ballistic'' action
change in contrast to the diffusive action changes of chaotic
trajectories.  The variance is therefore proportional to the square of the
period.  According to the numerical calculations discussed ahead,
ballistic action changes hold equally well for integrable and
near-integrable systems, but this is not true for mixed phase space
systems.  Thus, excluding mixed systems,
\begin{eqnarray}
      \label{eq:action_squared}
     &&  \left \langle \left\{ \left( \partial \overline{S}_{\bf M}
        \over \partial \lambda \right)^2
        [J_0^2(s_\lambda) + \tilde{a}^2 J_1^2(s_\lambda)] \right. \right. 
        \nonumber \\
     && +  \left. \left. \left( \partial
        \Delta S_{\bf M}
        \over \partial \lambda \right)^2
        \left[ {\tilde{a}^2 \over 4} (J_0(s_\lambda) - J_2(s_\lambda))^2
        + J_1^2(s_\lambda) \right] \right\}  \right \rangle_{\bf M} \nonumber
\\ && = \zeta(E, \lambda; \hbar) T^2
\end{eqnarray}
It was shown in~\cite{ullmo:96} that the expression involving Bessel
functions collapsed to the usual Berry-Tabor expression for most tori
(with a minor adjustment for the slight distortion of the torus,
i.e.~$s_\lambda = 0,\partial \Delta S_{\bf M}/ \partial \lambda = 0$).
Only very slowly as $\hbar \rightarrow 0$, did tori other than those
possessing the lowest order ${\bf M}$ require the Bessel function
weightings.  For these higher order tori, Eq.~(\ref{eq:action_squared})
is understood as being just the mean square of $\partial \overline{S}_{\bf
M} / \partial \lambda$.

Performing the integral, the variance of the level velocities becomes
simply
\begin{equation}
      \label{eq:integrable_lv_variance}
      \sigma^2_E \approx \zeta(E, \lambda; \hbar)
\end{equation}
which is only very weakly dependent on $\hbar$ for those low order tori
requiring the Bessel function weightings.  Thus, in the integrable
limiting case, the $\hbar$-dependence disappears altogether.

For mixed phase space systems, the question is whether a generalization of
Eqs.~(\ref{eq:sum_rule},\ref{eq:action_squared}) is possible; for the
fully chaotic case, the expressions are modified by replacing $T^2$ by
$T$ and the coefficients change.  Is it possible to have fractionally
weighted parts of both expressions?  First, instead of summing over
rational tori $\bf M$, the sum is considered over all periodic orbits.
Given the diffusive action change behavior for fully chaotic systems and
ballistic for regular systems, conceptually, the full set of action
changes might decompose into a linear and a quadratic component.
Continuing with this logic, Eq.~(\ref{eq:sum_rule}) would apply for the
periodic orbits with ballistic action changes and be weighted by the
appropriate fraction of phase space volume.  It is important to note a
couple of subtleties here.  Both a stable and unstable orbit combined to
define $\{ \overline{S}_{\bf M},\Delta S_{\bf M} \}$.  Thus, the quadratic
and linear components cannot be as simple as treating the stable orbits
ballistically and the unstable orbits diffusively.  Perhaps, at least as
many orbits with positive Lyapunov exponent as stable orbits should be
treated ballistically.  Secondly, the key ingredient, $\overline{S}_{\bf
M}$, is the average of said stable and unstable orbit actions.  Taking
the mean square of the sum of two orbits's actions is equal to the square
of the sum only if the action changes are nearly equal.  Excluding the
lowest order tori, this does turn out to be roughly true.

Similarly, the remaining unstable orbits would be treated with
the appropriate (chaotic) versions of
Eqs.~(\ref{eq:sum_rule},\ref{eq:action_squared}).  With this
separation, the proposition would be that the level velocity variance
is the sum of the near-integrable result and the
previously derived chaotic result~\cite{cerruti:01},
\begin{equation}
      \label{eq:lv_variance_mixed}
      \sigma^2_E \approx \zeta(E, \lambda; \hbar) f
       + {g K(E, \lambda) \hbar^{d - 1} \over \pi V} (1 - f)
\end{equation}
where $f$ is the fraction of ballistically-behaving orbits, and $g$ is a
symmetry factor for the system.  The quantities $\zeta(E, \lambda; \hbar)$
and $K(E, \lambda)$ would be determined by the ballistic and diffusive
versions of Eq.~(\ref{eq:action_squared}), respectively.  The section on
the kicked rotor will address this issue.

\subsection{Overlap intensities}
\label{oi}

Next, the overlap intensities are investigated and a semiclassical
derivation of their variance is obtained.  For an integrable system the
overlap intensities are best expressed in action-angle coordinates;
details are in appendix~\ref{intensities} and the result is
\begin{equation}
      \label{eq:var_intensity}
      \sigma^2_\alpha \approx {\prod \sigma^2_j \over \pi^d}
      \left \langle \exp \left[ {-2 (\Delta {\bf I})^2
      \over \hbar^2}\right] \right \rangle_{\bf M}
\end{equation}
where in the exponential argument of the wave packet, each component is
$(\Delta {\bf I})_j \equiv \sigma_j ({\bf I} - {\bf I}_\alpha)_j$.  The
variance is a Gaussian weighted average of the action of the wave packet
over the rational tori.

The Gaussian weighting can be evaluated by replacing the $\bf{M}$
averaging by an average of the action variables over the energy surface
\begin{eqnarray}
      \left \langle \exp \left[ {-2 (\Delta {\bf I})^2
        \over \hbar^2}\right] \right \rangle_{\bf M}
        &\approx& {(2\pi)^d \over V} \int \exp \left[ {-2 (\Delta {\bf I})^2
        / \hbar^2}\right] \nonumber \\
      && \times \delta(E - H_0({\bf I})) d{\bf I}
      \end{eqnarray}
The Hamiltonian in the numerator is expanded about ${\bf I}_\alpha$ giving
\begin{equation}
      H_0({\bf I}) = H_0({\bf I_\alpha})
      + {\bm \omega}_\alpha \cdot ({\bf I} - {\bf I}_\alpha) + \cdots
\end{equation}
where $H_0({\bf I_\alpha}) = E$ and ${\bm \omega}_\alpha
= \partial H_0({\bf I_\alpha}) / \partial {\bf I}$.  Only linear terms
are necessary in the semiclassical limit as long as each width component
is chosen such that $\sigma_j/\hbar \rightarrow \infty$ as $\hbar
\rightarrow 0$; we take $\sigma_j \propto \hbar^{1/2}$ so that
uncertainty in ${\bf I}$ and ${\bf \theta}$ shrink similarly as $\hbar$
shrinks.  After performing the integration of the weighting over the
energy shell, the result is
\begin{eqnarray}
      \label{eq:dampening}
      \left \langle \exp \left[ {-2 (\Delta {\bf I})^2
        \over \hbar^2}\right] \right \rangle_{\bf M}
        &\approx& \left({\pi \hbar^2 \over 2}\right)^{(d - 1) / 2}
	\nonumber \\
      &\times& {1 \over V \prod \sigma_j \sqrt{\sum
\left({\omega_{\alpha j} / \sigma_j}\right)^2}}
\end{eqnarray}
which has an $\hbar$-scaling of $\hbar^{(d - 1) / 2}$.  Therefore, the
leading order $\hbar$-scaling for the intensity variance for
near-integrable systems is
\begin{equation}
      \sigma^2_\alpha =  {\prod \sigma_j \over V  \sqrt{\sum
\left({\omega_{\alpha j} / \sigma_j}\right)^2}}
     \propto \hbar^{(3d - 1) / 2}
\end{equation}

\subsection{Correlations}
\label{correlations}

Finally, the semiclassical theories for the overlap intensities and the
level velocities outlined above can now be used to construct the
correlation function.  Beginning with the covariance, it can be shown that
\begin{eqnarray}
      Cov_\alpha(\lambda) &=& \left \langle p_{\alpha n}(\lambda)
        {\partial E_n(\lambda) \over \partial \lambda} \right \rangle_n
        \nonumber \\
      &=& {2 \pi h^d \epsilon \over V}
        \left \langle S_\alpha(E; \lambda)
        {\partial N(E; \lambda) \over \partial \lambda} \right \rangle_E
\end{eqnarray}
The semiclassical expressions for the strength function and the parametric
derivative of the staircase function give
\begin{eqnarray}
      && Cov_\alpha(\lambda) = {-\epsilon 2^{(d + 3)/2} 
        \prod \sigma_j \pi^{1/2}
        \over V \hbar} \nonumber \\
      &\times& \left \langle \sum_{\bf M} {}^\prime |A_{\bf M}|
        |D_s|^{1/2} \left[ \left( \partial \overline{S}_{\bf M} \over
        \partial \lambda \right)
        [J_0(s_\lambda) - i\tilde{a}(\lambda)J_1(s_\lambda)]
        \right. \right.  \nonumber \\
      &+& \left. i \left( \partial \Delta S_{\bf M}
        \over \partial \lambda \right)
        \left(J_1(s_\lambda) + {i\tilde{a}(\lambda) \over 2}
        [J_0(s_\lambda) - J_2(s_\lambda)]\right)\right] \nonumber \\
      &\times& \left. 
        \exp \left( - {(\Delta {\bf I})^2 \over \hbar^2}-{2 \epsilon T_{\bf
        M} \over \hbar} \right) \right \rangle_E
\end{eqnarray}
Again, the diagonal approximation has been made in the above expression.
Following arguments parallel to those in deriving the variances leads to
the expression
\begin{eqnarray}
      Cov_\alpha(\lambda) &=& {-2 \prod \sigma_j \over \pi^{d/2} \tau_H}
        \int {dT \over T} e^{-2 T / \tau_H} \nonumber \\
       && \times
        \left \langle \left( {\partial S_{\bf M} \over \partial \lambda} 
        \right)
        \exp \left[ {-(\Delta {\bf I})^2 \over \hbar^2}\right] \right
        \rangle_{\bf M}
\end{eqnarray}
for integrable systems where $\tau_H = \hbar / \epsilon$ is the
Heisenberg time.

In order to make further progress in understanding the correlation
function,  it is necessary to know the time dependence of the ${\bf
M}$-averaged  expression.  It is important to note that the Gaussian
weighting factor  will not decouple from the parametric action derivative
at long times and thus, the local average of the derivatives of the
action is not necessarily zero.
The weighted action derivatives can be approximated by the same method as
used in the calculation the variance of the level velocities, so
for integrable systems
\begin{eqnarray}
      \label{eq:action_intensity_corr}
      && \left \langle \left( {\partial S_{\bf M} \over \partial \lambda} 
        \right)
        \exp \left[ {-(\Delta {\bf I})^2 \over \hbar^2}\right]
        \right \rangle_{\bf M} \nonumber \\
      && \approx {\int (\partial S_{\bf M} / \partial \lambda)
        \exp[ -(\Delta {\bf I})^2 / \hbar^2] \delta(T - T_{\bf M}) d{\bf M}
        \over \int \delta(T - T_{\bf M}) d{\bf M}} \nonumber \\
      && \approx T \xi(\lambda)
\end{eqnarray}
Note that the above average is linear with the period.
The $\hbar$-scaling of $\xi(\lambda)$ is the same as for the
average Gaussian weighting in the previous subsection which is
$\hbar^{(d - 1) / 2}$.

The final result of the covariance is
\begin{equation}
      \label{eq:covariance}
      Cov_\alpha(\lambda) \approx {-\xi(\lambda) \prod \sigma_j \over \pi^{d/2}}
\end{equation}
Combining this with the results above for the variances generates the
semiclassical expression for the correlation function
\begin{equation}
      \label{eq:correlation}
      {\cal C}_\alpha(\lambda) \approx {-\xi(\lambda)
      \over \zeta^{1/2}(E, \lambda)
      \left \langle \exp \left[ -2 (\Delta {\bf I})^2
      / \hbar^2 \right] \right \rangle^{1/2}_{\bf M}}
\end{equation}
where the expectation value is evaluated in Eq.~(\ref{eq:dampening}).  For
near-integrable systems in the semiclassical limit, the above result
implies  that
$Cov_\alpha(\lambda)
\propto
\hbar^{(2d - 1)/2}$ whereas the correlation ${\cal C}_\alpha(\lambda)
\propto
\hbar^{(d - 1)/4}$ to leading order.

\section{Results}

\subsection{Rectangular and box billiards}

First, we consider billiards that are totally separable in Cartesian
coordinates and thus completely integrable.  These types of billiards
have been used in the study of spectral rigidity~\cite{berry:85}.
In particular, we examine a rectangular billiard with sides $a, b$ and
a box billiard with sides $a, a, b$; the sidelength $a$ is
varied.  The density of states is kept constant by not changing
the area ${\cal A}$ of the rectangle or the volume ${\cal V}$ of the box.  The
action/angle variables are proportional to the momentum/position
coordinates in each direction,
\begin{equation}
      I_j= \ell_j p_j / \pi \ \ \ \mbox{and} \ \ \ \theta_j = \pi q_j / \ell_j
\end{equation}
where $\ell_j$ is the side length of the $j$th side.  The Hamiltonians are
given by
\begin{equation}
      H({\bf I}) = {\pi^2 \over 2 m} \sum_j \left( {I^2_j \over 
\ell^2_j} \right)
\end{equation}

\begin{figure}[!t]
      \begin{center}
      \epsfig{file=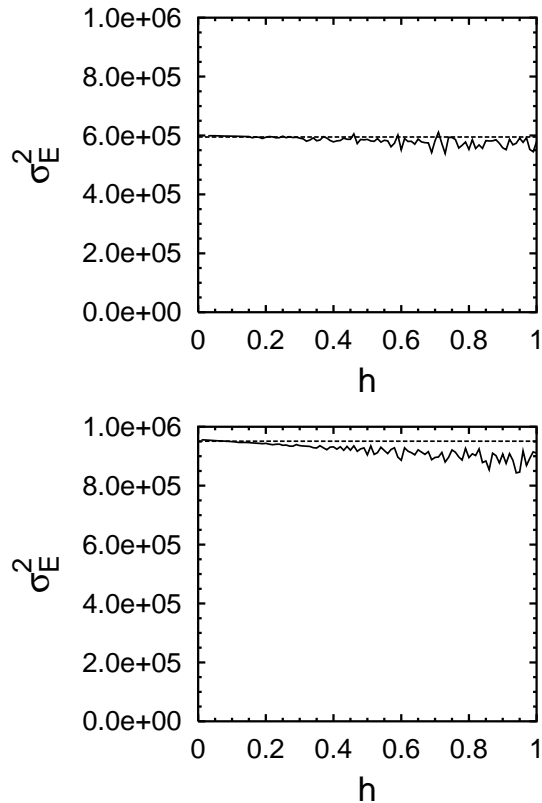, width=3in}
      \end{center}
      \caption{Variance of the level velocities for rectangular and box
      billiards.  The area and volume are taken to be unity, while the
      sidelength $a = 1.1$; the energy is 600.  The upper panel is
      for the rectangle and the lower panel is for the box.  The solid
      line is the quantum results while the dashed line is the
      semiclassical theory.}
      \label{fig:lv_billiard}
\end{figure}

The classical action of the rectangle for a closed orbit on a given resonant
torus ${\bf M}$ is
\begin{equation}
      S_{\bf M} = \sqrt{8mE} [M^2_1 a^2 + M^2_2 ({\cal A} / a)^2]^{1/2}
\end{equation}
the period is
\begin{equation}
      T_{\bf M} = \sqrt{2m [M^2_1 a^2 + M^2_2 ({\cal A} / a)^2] / E}
\end{equation}
and the derivative of the action along the orbit is
\begin{equation}
      {\partial S_{\bf M} \over \partial a}
      = {\sqrt{8mE} \over a}
      {M^2_1 a^2 - M^2_2 ({\cal A} / a)^2
      \over [M^2_1 a^2 + M^2_2 ({\cal A} / a)^2]^{1/2}}
\end{equation}
Similarly, the classical action of the box is
\begin{equation}
      S_{\bf M} = \sqrt{8mE} [(M^2_1 + M^2_2) a^2
      + M^2_3 ({\cal V} / a^2)^2]^{1/2}
\end{equation}
the period is
\begin{equation}
      T_{\bf M} = \sqrt{2m [M^2_1 a^2 + M^2_2 a^2 + M^2_3 ({\cal V} /
a^2)^2] / E}
\end{equation}
and the derivative of the action is
\begin{equation}
      {\partial S_{\bf M} \over \partial a}
      = {\sqrt{8mE} \over a}
      {(M^2_1 + M^2_2) a^2 - 2 M^2_3 ({\cal V} / a^2)^2
      \over [(M^2_1 + M^2_2) a^2 + M^2_3 ({\cal V} / a^2)^2]^{1/2}}
\end{equation}
Using the above quantities and performing the integration in
Eq.~(\ref{eq:action_squared}), the variance of the action derivatives
are $2 E^2 T^2 / a^2$ and $16 E^2 T^2 / 5 a^2$ for the rectangle and
the box, respectively.  Hence, the respective values of $\zeta(E, a)$ are
$2 E^2 / a^2$ and $16 E^2 / 5 a^2$, which approach the
variances of the level velocities in the semiclassical limit.
Figure~\ref{fig:lv_billiard} demonstrates how well $\zeta(E, a)$
approximates the level velocities especially as $\hbar \rightarrow 0$.
Note that the level velocities are virtually independent of $\hbar$
as predicted.

\begin{figure}[!t]
      \begin{center}
      \epsfig{file=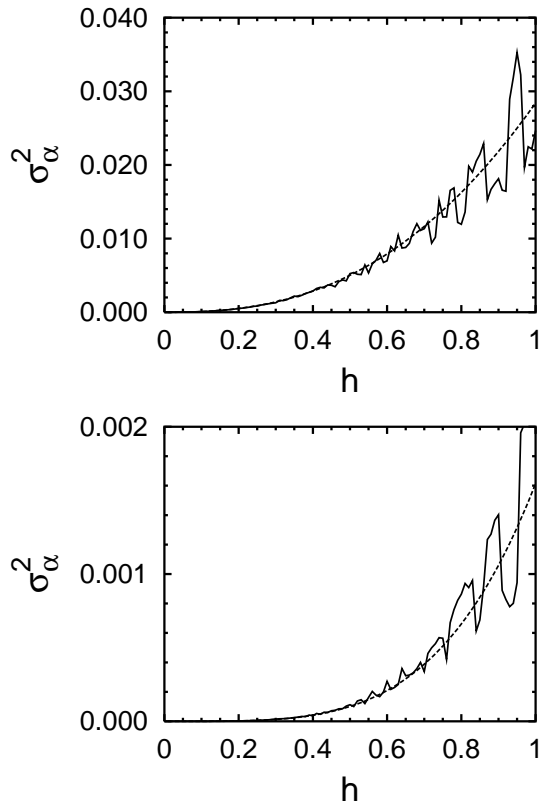, width=3in}
      \end{center}
       \caption{Variance of the intensities for the rectangular and box  
      billiards.  The area and volume are taken to be unity, while the
      sidelength $a = 1.1$; the energy is 600.
      The upper panel is for the rectangle and the lower panel is for the box.
      The solid line is the quantum results while the dashed line is the
      semiclassical theory.}
      \label{fig:intensity_billiard}
\end{figure}

The $\hbar$-scaling of the intensity variance is predicted to be
$\hbar^{5/2}$ and $\hbar^4$, respectively, for the rectangular and box
billiards.  Using Eq.~(\ref{eq:dampening}) with $\omega_{\alpha, j}
= \pi^2 I_{\alpha, j} / m \ell^2_j$ the semiclassical theory and quantum
intensity variances are compared in Fig.~\ref{fig:intensity_billiard}.
The agreement is quite good as $\hbar \rightarrow 0$.

\begin{figure}[!t]
      \begin{center}
      \epsfig{file=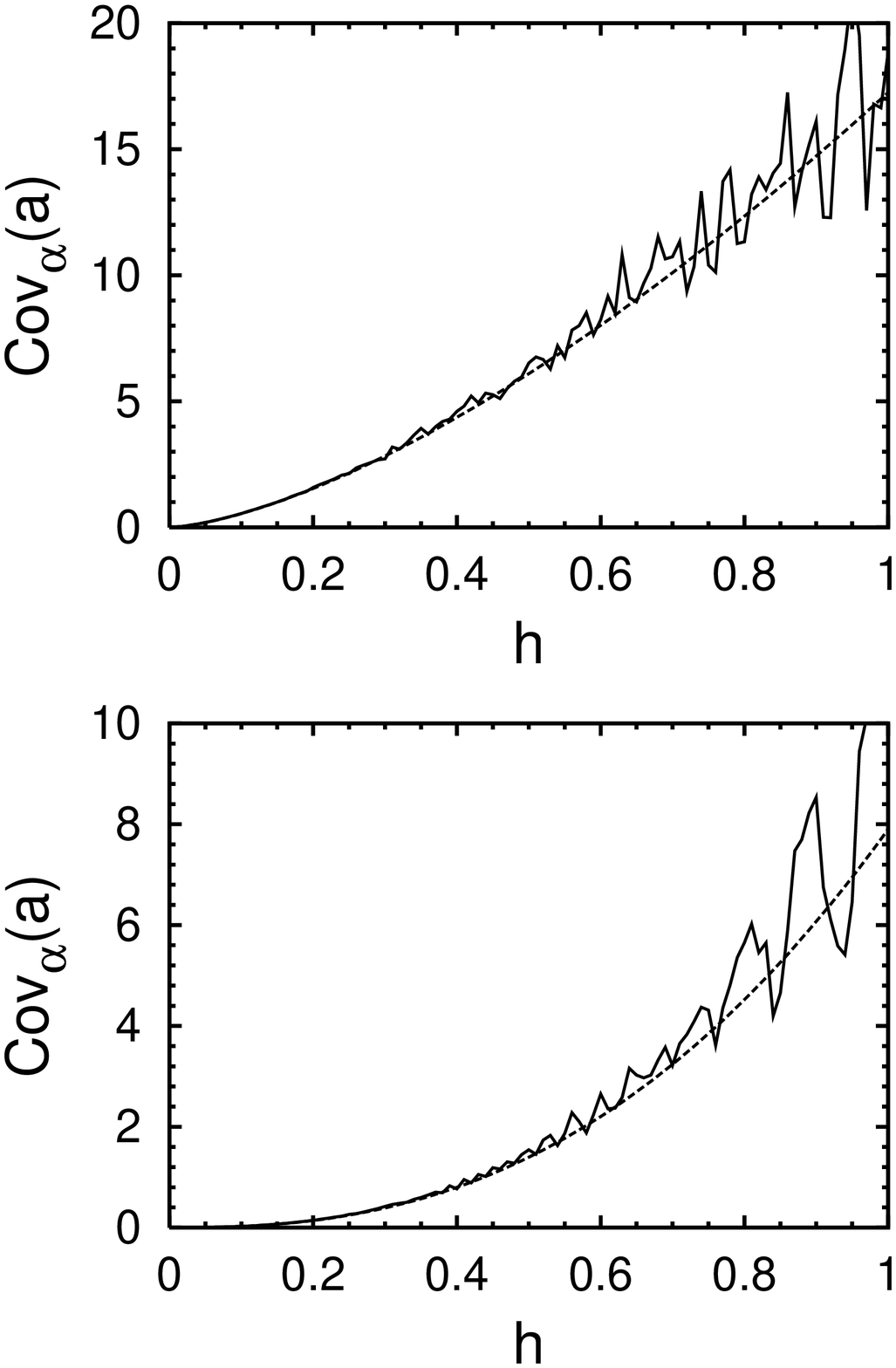, width=3in}
      \end{center}
      \caption{Covariance for the rectangular and box billiards.
      The area and volume are taken to be unity, while the
      sidelength $a = 1.1$; the energy is 600.
      The upper panel is for the rectangle and the lower panel is for the box.
      The solid line is the quantum results while the dashed line is the
      semiclassical theory.}
      \label{fig:covariance_billiard}
\end{figure}

\begin{figure}[!t]
      \begin{center}
      \epsfig{file=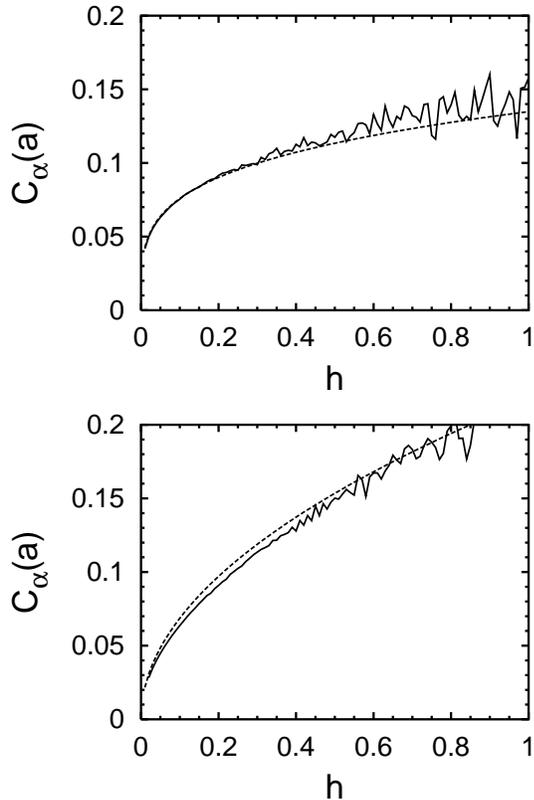, width=3in}
      \end{center}
      \caption{Correlation for the rectangular and box billiards.
      The area and volume are taken to be unity, while the
      sidelength $a = 1.1$; the energy is 600.
      The upper panel is for the rectangle and the lower panel is for the box.
      The solid line is the quantum results while the dashed line is the
      semiclassical theory.}
      \label{fig:correlation_billiard}
\end{figure}

The last two quantities, the covariance and correlation, are shown in
Fig.~\ref{fig:covariance_billiard} and
Fig.~\ref{fig:correlation_billiard}, respectively, along with the results
from the semiclassical theory.  The  values of $\xi(E, a)$ are obtained
from Eq.~(\ref{eq:action_intensity_corr}). For the rectangle
\begin{equation}
      \xi(E, a) = {2 \pi^{3/2} \hbar \over a m \sigma \sqrt{2 m E}}
      \left( {I^2_{\alpha, 1} \over a^2}
      - {I^2_{\alpha, 2} \over ({\cal A} / a)^2} \right)
\end{equation}
and for the box
\begin{equation}
      \xi(E, a) = {\pi^2 \hbar^2 \over a m^2 E \sigma^2}
      \left( {I^2_{\alpha, 1} \over a^2} + {I^2_{\alpha, 2} \over a^2}
      - {2 I^2_{\alpha, 3} \over ({\cal V} / a^2)^2} \right)
\end{equation}
The agreement is excellent in both cases.  Note that we have used no
fitting parameters in any of the figures in these subsection.

\subsection{Standard map}

In this subsection we compare the semiclassical theories to the numerical
results for the standard map.  The classical standard map is defined by
\begin{eqnarray}
      q_{i+1} &=& q_i + p_{i+1} \ \ \mathrm{mod} (1) \nonumber \\
      p_{i+1} &=& p_i - (k / 2 \pi) \sin(2 \pi q_i) \ \ \mathrm{mod} (1)
\end{eqnarray}
where the ``kicking strength'' $k$ is the varied parameter.
The map displays the entire range of classical dynamics from
integrable at $k = 0$ to fully chaotic beyond $k \approx 5$.

Instead of the velocities of the eigenvalues of the Hamiltonian, it is
the velocities of the eigenangles for the standard map which are of
interest.  The matrix elements of the quantum map with
$N$ discrete levels can be written~\cite{al:97}
\begin{eqnarray}
      \langle n | U | n^\prime \rangle &=& {1 \over \sqrt{iN}}
        \exp[i \pi (n - n^\prime)^2/N] \nonumber \\
      && \times \exp \left( i {kN \over 2 \pi}
        \cos[2 \pi (n + a) / N]\right)
\end{eqnarray}
where $n, n^\prime = 0, \dots, N-1$; $a$ is a phase term which we set
equal to zero.  The effective Planck constant is given by $h = 1/N$.  The
eigenvalues of the propagator are $U|\psi_j\rangle
= \exp(-i\phi_j)|\psi_j\rangle$ with real eigenangles $\phi_j$.
The ``level velocities'' are given by the Hellmann-Feynman theorem
\begin{equation}
      {d \phi_j \over dk} = {N \over 2\pi}
      \langle \psi_j | \cos(2 \pi q) | \psi_j\rangle
\end{equation}

The smoothed density of states for the eigenangles~\cite{oconnor:91} is
\begin{equation}
      d_\epsilon(\phi) = {N \over 2 \pi} +
      {1 \over 2 \pi} \sum^\infty_{n = -\infty}
      \exp(in\phi) \mathrm{Tr} \, U^n \exp(-n \epsilon)
\end{equation}
where the sum excludes $n = 0$ and the trace of the propagator in the
integrable regime is
\begin{equation}
      \mathrm{Tr} \, U^n = \sum_{\bf M} \sqrt{N \over in}
      \exp(2 \pi i N S_{\bf M})
\end{equation}
Integrating over the angle, the smoothed spectral staircase is
\begin{eqnarray}
      N_\epsilon(\phi) &=& {N \phi \over 2 \pi} -
        {1 \over 2 \pi} \sum^\infty_{n = -\infty} \sum_{\bf M}
        {1 \over n} \sqrt{i N \over n} \exp(in\phi) \nonumber \\
      && \times \exp(2 \pi i N S_{\bf M}) \exp(-n \epsilon)
\end{eqnarray}
The resonant tori ${\bf M}$ including repetitions for the standard map
are given by a constant momentum equal to the rational fraction $m / n$
where $m = 0, \dots, n - 1$ and $n$ is the period of the torus.  The
derivative of the action with respect to the kicking strength is
\begin{equation}
      {\partial S_{\bf M} \over \partial k}
      = {1 \over (2 \pi)^2} \sum^n_{i = 1} \cos(2 \pi q_i)
\end{equation}
For the integrable case ($k = 0$), the periodic points are
$q_i = q_0 + m \cdot i / n \ \ \mathrm{mod} (1)$
and hence, $\partial S_{\bf M} / \partial k = 0$
for all resonant tori.  Thus, by Eqs.~(\ref{eq:action_squared})
and (\ref{eq:integrable_lv_variance}) the variance of the level
velocities is zero.

\begin{figure}[!t]
      \begin{center}
      \epsfig{file=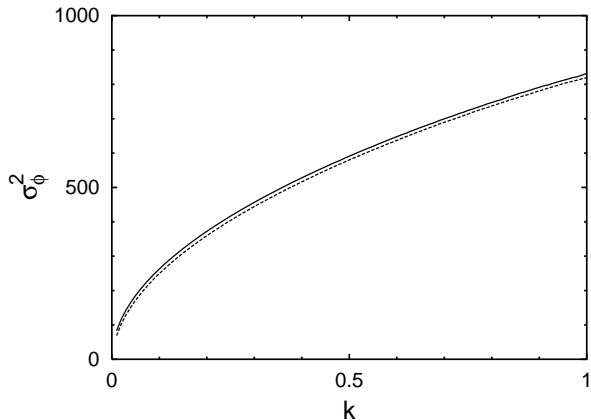, width=2.25in, angle=270}
      \end{center}
      \caption{The variance of the level velocities for the standard map
      ($N = 500$).  The solid line is the quantum results while the dashed
      line is the semiclassical theory using the near-integrable results
      of Eq.~(\ref{eq:standardmap_lv_variance}).}
      \label{fig:variance_lv}
\end{figure}

In the near-integrable regime, the variance of the eigenangles
is approximately given by the result
\begin{equation}
      \label{eq:standardmap_lv_variance}
      \sigma^2_\phi \approx 4 \pi^2 \zeta N^2
\end{equation}
Note that the eigenangles are divided by an effective $\hbar$
making the variance of the eigenangles quadratic in $N$.  We do not have
an analytic formula for $\zeta$, but it can be obtained numerically from
the quadratic period dependence of the variance of the action derivatives
averaged over the tori.  For $k=0$, it is straightforward to calculate
analytically the positions of the stable and unstable periodic orbits that
arise as soon as integrability is broken.  Curiously, using these points
as initial conditions for the periodic orbits throughout the
near-integrable regime, roughly $k \in [0,2]$, gives results within a few
percent of that found by locating the true periodic orbits.  In order to
make long time calculations, this compromise is indispensable and does
not harm the approximation noticeably.  In
Fig.~\ref{fig:variance_lv}, the near-integrable theory is shown to give
excellent agreement with the quantum eigenangle velocity variance even
though there exists a significant proportion of unstable orbits for the
larger values of $k$ shown that are being treated as though they were
stable.

In the mixed regime, the corresponding proposition of
Eq.~(\ref{eq:lv_variance_mixed}) is
\begin{equation}
      \label{eq:standardmap_lv_variance_mixed}
      \sigma^2_\phi \approx 4 \pi^2 \zeta(k) f N^2
       + g {D(k) \over 4 \pi^2} (1 - f) N
\end{equation}
where $g = 4$ and $D(k)$ is the classical action diffusion constant.
It was shown previously~\cite{lakshminarayan:99} that a direct method based
on classical orbits gave a good correspondence between the theory and
quantum values for
\begin{eqnarray}
      \label{eq:mixed}
      \sigma^2_\phi &\approx& {N^2 \over 4 \pi^2} \int^1_0 \int^1_0
        V(q_0) \langle V(q^{(n)}(q_0, p_0)) \rangle_n dp_0 dq_0 \nonumber \\
      &\approx& c_1(k) N + c_2(k) N^2
\end{eqnarray}
where $V(q) = \cos(2 \pi q)$ but no understanding was given of the
two coefficients.  Equation~(\ref{eq:standardmap_lv_variance_mixed}) gives
one plausible, but naive, interpretation.

In order to test numerically the separation of diffusive and ballistic
contributions in the mixed regime, it is impossible to work with the
periodic orbits.  They proliferate exponentially and must be found
numerically.  The time over which this can be done by brute force
calculation is too short for our purposes.  Instead, a Monte Carlo
technique is needed.  We have devised a method that begins with a uniform
sampling of initial conditions in the phase space, and relies on the
properties of finite time Lyapunov exponents~\cite{grassberger:88}.
Although, a stable periodic orbit has a vanishing Lyapunov exponent, at a
partial period of its motion, the trace, $Q$, of its stability matrix can
grow as a power of the propagation time.  This can give a stable orbit the
appearance of being unstable depending on circumstances.  For example,
constructing a surface of section based on whether $|Q|<2$ fails
miserably.  If instead, one seeks the least square deviation of the time
development of $Q$ from the form
\begin{equation}
\label{fit}
\bar Q = aT^\gamma\exp \mu T
\end{equation}
then the estimate of the Lyapunov exponent can be used to judge whether
an orbit is in the stable or unstable part of the phase space; i.e. the
construction of the correct surface of section is recovered in this way.

\begin{figure}[!t]
      \begin{center}
      \epsfig{file=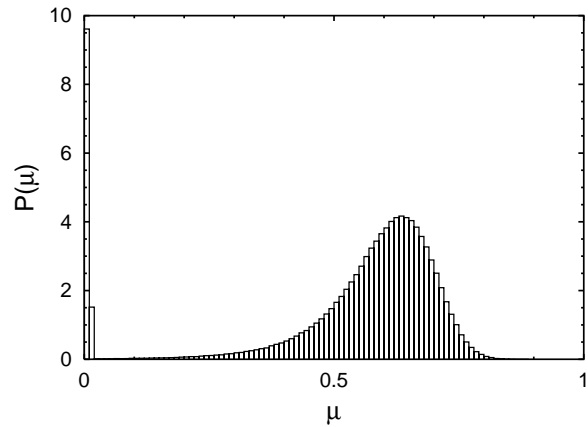, width=2.25in, angle=270}
      \end{center}
     \caption{Histogram of finite time Lyapunov exponents in the standard map
     in the mixed regime.  The parameters are $k = 2.7$ and $T = 500$.
     The values of $\mu$ were obtained by a least squares fit of the
     natural logarithm of Eq.~(\ref{fit}).}
     \label{fig:histogram}
\end{figure}

Figure~\ref{fig:histogram} gives a histogram of the finite time Lyapunov
exponents, $\mu$, calculated using Eq.~(\ref{fit}) for the orbits in this
regime.  The orbits are separated into three rough categories
depending upon their stabilities as mentioned in the previous
section.  The broadened $\delta$-function feature near zero represents the
stable orbits that have action derivatives with a quadratic dependence
upon time.  The set of most highly unstable or chaotic trajectories give
an approximate log-normal distribution~\cite{grassberger:88}.  In between
these two features lie the weakly unstable orbits.  It turns out that the
coefficient $\zeta(k)$ converges well if averaged over the stable orbits
and up to a similar number of the weakly unstable orbits.  However, it
also turns out that when compared to the coefficient $c_2(k)$ of
Eq.~(\ref{eq:mixed}), the fraction $f$ deduced by their comparison is
inconsistent with the relative fraction used to derive $\zeta(k)$ for $k$
values in the range $[1.5,4.0]$; see appendix \ref{table}.

A second inconsistency with the simplistic separation problem is that the
remaining strongly unstable orbits do not show a purely diffusive
behavior.  They show both a quadratic and linear dependence in their
action derivative variances; i.e. the numerical evaluation of $D(k)$ is
suspect.  It could also be evaluated according to its asymptotic
expression~\cite{lakshminarayan:99}, but again requires a large value of
$f$ to be consistent with the coefficient $c_1(k)$ unless $k$ exceeds roughly
$5$ or so.  At that point $f\approx 0$ and the system is basically in the
fully chaotic regime.  We attempted several schemes of separating out
orbits according to ballistic or diffusive behaviors, none of which
worked and are outlined in appendix \ref{table}.

\begin{figure}[!t]
      \begin{center}
      \epsfig{file=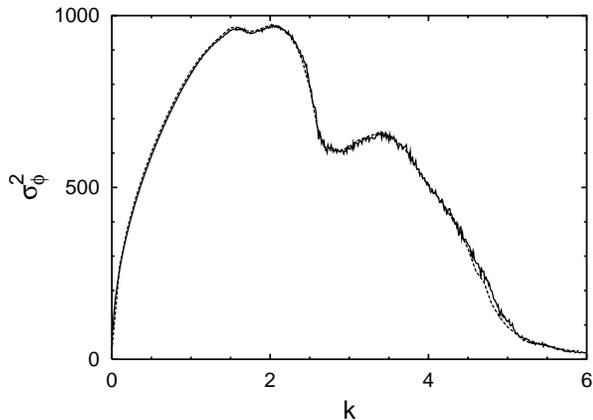, width=2.25in, angle=270}
      \end{center}
      \caption{The variance of the level velocities for the standard map
      ($N = 500$).  The solid line is the quantum results while the dashed
      line the classical phase space average of Eq.~(\ref{eq:mixed}).  The
      time, $n$, is taken to be $N/g=125$.}
      \label{fig:variance_lv_arul}
\end{figure}

In spite of the lack of apparent separability, the simplistic expression,
Eq.~(\ref{eq:mixed}) works quite well.  We recalculate the results
of~\cite{lakshminarayan:99} taking into account the Heisenberg time
associated with the mean eigenangle spacing in
Fig.~\ref{fig:variance_lv_arul}.  As a function of kicking strength
$k$ the agreement between the semiclassical theory and the variance of the
eigenangle velocities is quite good throughout the entire transition from
integrability to full chaos.

Next we turn to the overlap intensities.  Since the energy is not
conserved for maps, Eq.~(\ref{eq:dampening}) does not apply.  The variance
is obtained directly from Eq.~(\ref{eq:var_intensity}) where the average
is over all resonant tori; it becomes just the Gaussian normalization
factor.  The width of the Gaussian wave packet in Cartesian coordinates is
taken to be proportional to the square root of
$\hbar$, $\sigma = \beta \sqrt{\hbar} = \beta / \sqrt{2 \pi N}$.  Thus,
for $k = 0$ the variance of the intensities is
\begin{equation}
      \label{eq:intensity_standardmap}
      \sigma^2_\alpha \approx \sqrt{\beta \over N^3}
\end{equation}
This expression does not take into account the reflection, $(q, p)
\rightarrow (1 - q, 1-p)$, and time reversal symmetries of the kicked
rotor~\cite{dittrich:91}.  Thus, we symmetrize the wave packet
accordingly.

\begin{figure}[!t]
      \begin{center}
      \epsfig{file=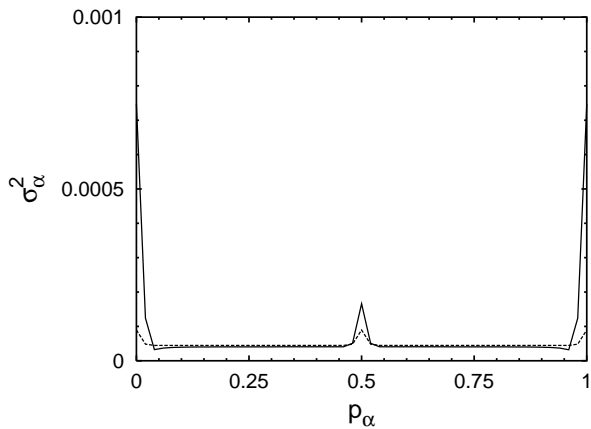, width=2.25in, angle=270}
      \end{center}
      \caption{Variance of the intensities of the standard map for a section of
      phase space at $q_\alpha = 0.25$.  The parameters are $N = 500$ and
      $k = 0.01$. The solid line is the quantum results while the dashed line
      is the semiclassical theory.}
      \label{fig:intensity_standardmap}
\end{figure}

Figure~\ref{fig:intensity_standardmap} shows the good agreement between
the semiclassical theory and the intensity variance for the integrable
regime at $k = 0.01$.  The only exceptional cases are the tori at $p = 0$
and $p = 0.5$ which happen to be self-symmetric and on the lowest order
resonances.  Whereas for $k=0$, they would lead to double the
intensity variance and be accurately accounted for once the symmetry
was fully incorporated, at $k=0.01$ they are strongly affected by their
location on low order resonances.  As we do not have the generalizations
necessary for those cases, we avoid the resonance zones instead.

\begin{figure}[!t]
      \begin{center}
      \epsfig{file=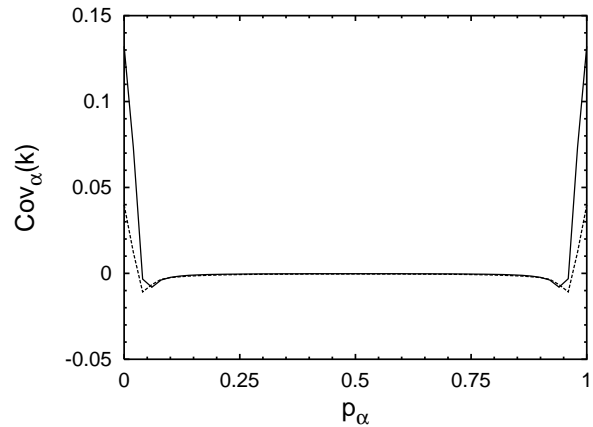, width=2.25in, angle=270}
      \end{center}
      \caption{Covariance of the standard map for a section of
      phase space at $q_\alpha = 0.25$.  The parameters are $N = 500$ and
      $k = 0.01$.  The solid line is the quantum results while the dashed line
      is the semiclassical theory.}
      \label{fig:covariance_standardmap}
\end{figure}

\begin{figure}[!t]
      \begin{center}
      \epsfig{file=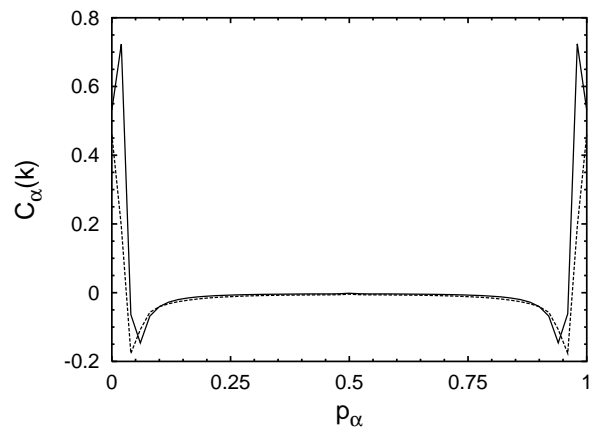, width=2.25in, angle=270}
      \end{center}
      \caption{Correlation of the standard map for a section of
      phase space at $q_\alpha = 0.25$.  The parameters are $N = 500$ and
      $k = 0.01$.  The solid line is the quantum results while the dashed line
      is the semiclassical theory.}
      \label{fig:correlation_standardmap}
\end{figure}

Finally, we turn to the correlations.  The covariance between the
intensities and level velocities is given by Eq.~(\ref{eq:covariance})
divided by $\hbar = 1 / 2 \pi N$ since the level velocities in the
standard map are actually eigenangle velocities.  After symmetrizing the
wave packet, Fig.~\ref{fig:covariance_standardmap} compares the
semiclassical prediction with the quantum results.  Also, the correlation
is compared in Fig.~\ref{fig:correlation_standardmap}. In both cases the
agreement between the semiclassical theory and the quantum results is
good.

\begin{figure}[!t]
      \begin{center}
      \epsfig{file=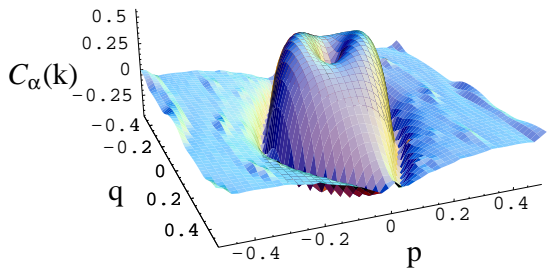, width=1.5in, angle=270}
      \end{center}
      \caption{Correlation of the standard map in the near-integrable regime.
      The parameters are $N = 500$ and $k = 0.5$.}
      \label{fig:correlation_near_standardmap}
\end{figure}

In the near-integrable regime, the covariance and correlation is strongest
for resonance zones; see Fig.~\ref{fig:correlation_near_standardmap} and
Ref.~\cite{keshavamurthy:02}.  The quantizing tori are deforming at a
greater rate inside the resonance while the tori outside the resonance
are only slightly perturbed.  Thus, the associated level velocities are
larger for this area of phase space.  Also, note that there is a dip in
the correlation near the stable periodic orbit at the center of the
resonance zone.  The dip is only present in the correlation and not in
the covariance. The correlation is divided by the variance of the
intensities which is large near the center of the resonance.  The
intensity variance is divided by the winding numbers,
Eq.~(\ref{eq:dampening}), which are small near the center of the resonance
making the intensity variance large.

\section{Discussion}

We have derived semiclassical expressions of the correlation of level
velocities and overlap intensities in near-integrable systems.  Their
$\hbar$-scaling is very different than previously derived in the chaotic
limit~\cite{cerruti:01} where the level velocity variance is
\begin{equation}
      \sigma^2_E \propto \hbar^{d - 1}
\end{equation}
and the intensity variance~\cite{note} is
\begin{equation}
      \sigma^2_\alpha \propto \hbar^{2d - 1}
\end{equation}
In addition, for chaotic systems the action derivatives will eventually
decouple from the weightings in the correlation so that in the semiclassical
limit the random matrix theory prediction is recovered stating that no
correlations exist~\cite{tomsovic:96,cerruti:01}.  On the other hand,
for finite $\hbar$ with the average computed up to the Heisenberg time,
there is a possibility of short orbits that do not decouple
in chaotic systems, so correlations  may exist.  For near-integrable
systems there is no decoupling even in the long time limit.

Combining the previously derived results in chaotic systems with the
newly derived results for near-integrable systems, for mixed phase space
systems we attempted to separate the orbits for applying the Hannay-Ozorio
de Almeida sum rule to the different regions of phase space.  Several
criteria were chosen as a basis for the separation.  All of them led to
inconsistencies in the interpretation.  It seems that all of the unstable
orbits spend some part of their evolution mimicking stable motion and do
not give standard diffusion contributions to action derivatives.
Nevertheless, a different semiclassical
procedure~\cite{lakshminarayan:99}, not going through the Hannay-Ozorio
sum rule derivation, gives a theoretical prediction that matches well with
the level velocity variances independently of the dynamical regime of the
system.

Finally, the correlation between the level velocities and the intensities
has been shown to highlight resonance zones in near-integrable systems.
The resonance zones exhibit greater localization since the motion is
bounded within the separatrix.  Investigating the correlation and
covariance separably may yield more information about the system.
An example is the dip in the correlation that occurs
near the center of the resonance zone due to the stable periodic orbit;
the dip is not present in the covariance.

\acknowledgments

We gratefully acknowledge support from ONR Grant No.~N00014-98-1-0079 and
NSF Grant No.~PHY-0098027.

\appendix

\section{Level Velocities}
\label{lv}

The method~\cite{berry:94} to obtain the variance of the level velocities
begins with the smoothed spectral staircase
\begin{equation}
      N_\epsilon(E, \lambda) = \sum_n \theta(E - E_n(\lambda))
\end{equation}
Taking the derivative with respect to the parameter, we obtain
\begin{equation}
      {\partial N_\epsilon(E, \lambda) \over \partial \lambda}
      = \sum_n \delta(E - E_n(\lambda))
      {\partial E_n(\lambda) \over \partial \lambda}
\end{equation}
The quantity $\epsilon$ is an energy smoothing term which will be taken
smaller than the mean level spacing.  Upon energy averaging
\begin{equation}
      \left\langle {\partial N_\epsilon(E, \lambda) \over \partial \lambda}
      \right\rangle_E
      = \overline{d}(E, \lambda)
      \left\langle {\partial E_n(\lambda) \over \partial \lambda}
\right\rangle_n
\end{equation}
where $\overline{d}(E, \lambda)$ is the mean level density which is the
reciprocal of the mean level spacing and proportional to $\hbar^{-d}$.
The mean level spacing is related to the phase space volume by
$V = \overline{d} \hbar^d$.  Thus, in order to obtain information about
the level velocities, we will evaluate the spectral staircase.

The semiclassical construction of the spectral staircase is broken into an
average and an oscillating part
\begin{equation}
      N_\epsilon(E, \lambda) = \overline{N}(E, \lambda) + N_{osc}(E, \lambda)
\end{equation}
The average staircase $\overline{N}(E, \lambda)$ is the Weyl term and to
leading order in $\hbar$ is given by
\begin{equation}
      \overline{N}(E, \lambda) = {1 \over h^d}
      \int \theta(E - H({\bf I}, {\bm \theta}; \lambda) d{\bf I} d{\bm \theta}
\end{equation}
The oscillating part of the spectral staircase is a sum over rational
tori with topology ${\bf M}$ of the unperturbed Hamiltonian,
$H_0({\bf I})$~\cite{ullmo:96}
\begin{eqnarray}
      N_{osc}(E, \lambda) &=& {2 \over \hbar^{(d - 1) / 2}}
        \sum_{\bf M} {}^{\prime} A_{\bf M} \nonumber \\
        && \times \mathrm{Re} \left\{\exp\left({i \overline{S}_{\bf M}(\lambda)
        \over \hbar}
        - {i \eta_{\bf M} \pi \over 2} + {i \beta_{\bf M} \pi \over 4}\right)
        \right. \nonumber \\
      && \times [J_0(\Delta S_{\bf M}(\lambda)/ \hbar)
        - i\tilde{a}(\lambda)J_1(\Delta S_{\bf M}(\lambda) / \hbar)]
        \nonumber \\
      && \times \left. 
        \exp\left({-\epsilon T_{\bf M} \over \hbar} \right)\right\}
\end{eqnarray}
The phase $\eta_{\bf M} = {\bf M} \cdot {\bm \eta}$ where ${\bm \eta}$
are the Maslov indices and $\beta_{\bf M}$ is equal to the sign of the
determinant of the curvature matrix.

The variance of the level velocities is obtained by taking the mean square
of the counting function derivatives
\begin{eqnarray}
      && \left\langle \left(
         {\partial N_\epsilon(E, \lambda) \over \partial \lambda}
         \right)^2 \right\rangle_E
         = \left\langle \sum_n \sum_m
         {\partial E_n(\lambda) \over \partial \lambda}
         {\partial E_m(\lambda) \over \partial \lambda} \right. \nonumber \\
      && \times \left.
         \delta_\epsilon(E - E_n(\lambda)) \delta_\epsilon(E - E_m(\lambda))
         \frac{}{} \right\rangle_E
\end{eqnarray}
For a nondegenerate spectrum, the summation is nonzero only if $n = m$
because of the product of the two delta functions.  Since Lorentzian
smoothing is applied, then
\begin{equation}
      \delta^2_\epsilon(x) = {1 \over 2 \pi \epsilon} \delta_{\epsilon/2}(x)
\end{equation}
for $\epsilon \ll \overline{d}^{-1}$.  Thus we have
\begin{equation}
      \left\langle \left(
      {\partial N_\epsilon(E, \lambda) \over \partial \lambda}
      \right)^2 \right\rangle_E
      = {\overline{d} \over 2 \pi \epsilon} \left\langle \left(
      {\partial E_n(\lambda) \over \partial \lambda}
      \right)^2 \right\rangle_n
\end{equation}

The summation over the rational tori is most sensitive to
the changing actions and periods because
of the associated rapidly oscillating phases, i.e., the division by
$\hbar$ in the exponential.  Since the energy smoothing term $\epsilon$ is
taken smaller than a mean level spacing, it scales at least by $\hbar^d$
and the derivatives of the period vanish as $\hbar \rightarrow 0$.  Thus,
only the derivatives of the actions are considered, and the oscillating part
of the staircase yields
\begin{widetext}
\begin{eqnarray}
      && {\partial N_{osc}(E) \over \partial \lambda}
        \approx {2 \over \hbar^{(d + 1) / 2}}
        \sum_{\bf M} {}^{\prime} A_{\bf M}
        \mathrm{Im} \left\{-\exp\left({i \overline{S}_{\bf M}(\lambda) 
        \over \hbar}
        - {i \eta_{\bf M} \pi \over 2} + {i \beta_{\bf M} \pi \over 4}\right)
        \right. \nonumber \\
      && \times \left. \left[ \left( \partial \overline{S}_{\bf M} \over
        \partial \lambda \right)
        [J_0(s_\lambda) - i\tilde{a}(\lambda)J_1(s_\lambda)]
        + i \left( \partial \Delta S_{\bf M}
        \over \partial \lambda \right)
        \left(J_1(s_\lambda) + {i\tilde{a}(\lambda) \over 2}
        [J_0(s_\lambda) - J_2(s_\lambda)]\right)\right]
        \exp\left({-\epsilon T_{\bf M} \over \hbar} \right) \right\}
\end{eqnarray}
\end{widetext}
After making the diagonal approximation and energy averaging, the
variance of the level velocities for a near-integrable system is
\begin{widetext}
\begin{eqnarray}
      \sigma^2_E
        &\approx& {4 \pi \epsilon \over V \hbar}
        \sum_{\bf M} {}^{\prime} \left \langle A_{\bf M}^2
        \left\{ \left( \partial \overline{S}_{\bf M}
        \over \partial \lambda \right)^2
        [J_0^2(s_\lambda) + \tilde{a}^2 J_1^2(s_\lambda)] \right. \right.
        + 2 \left( \partial \overline{S}_{\bf M}
        \over \partial \lambda \right)
        \left( \partial \Delta S_{\bf M} \over \partial \lambda \right)
        \left[ {\tilde{a} \over 2} J_0(s_\lambda) (J_0(s_\lambda) -
        J_2(s_\lambda))
        + \tilde{a} J_1^2(s_\lambda) \right] \nonumber \\
      && + \left. \left. \left( \partial \Delta S_{\bf M}
        \over \partial \lambda \right)^2
        \left[ {\tilde{a}^2 \over 4} (J_0(s_\lambda) - J_2(s_\lambda))^2
        + J_1^2(s_\lambda) \right] \right\}
        \exp \left( {-2 \epsilon T_{\bf M} \over \hbar}\right) \right \rangle_E
\end{eqnarray}
\end{widetext}
The middle term vanishes since $(\partial \overline{S}_{\bf M}
/ \partial \lambda)$ and $(\partial \Delta S_{\bf M} / \partial \lambda)$
are uncorrelated and both average to zero.

\section{Overlap Intensities}
\label{intensities}

The variance of the intensities is derived by examining
the oscillating part of the strength function
\begin{equation}
      \label{eq:semi_strength}
      S_{\alpha, osc}(E, \lambda) = {-1 \over \pi} \mathrm{Im}
      \int \langle \alpha | {\bm \theta} \rangle
      G({\bm \theta}, {\bm \theta}^\prime; E)
      \langle {\bm \theta}^\prime | \alpha \rangle
      d{\bm \theta} d{\bm \theta}^\prime
\end{equation}
where
\begin{eqnarray}
      G({\bm \theta}, {\bm \theta}^\prime; E) &=&
        {1 \over i \hbar (2 \pi i \hbar)^{(d - 1) / 2}}
        \sum_j |D_s|^{1/2} \nonumber \\
      && \times \exp[i S_j({\bm \theta}, 
        {\bm \theta}^\prime; E) / \hbar
        - i \nu_j^\prime \pi / 2]
\end{eqnarray}
is the semiclassical energy Green's function.  The above sum is over all
paths that connect ${\bm \theta}$ to ${\bm \theta}^\prime$ on a given
energy surface $E$.  $D_s$ is a determinate involving second derivatives
of the actions
\begin{equation}
      D_s = \left|
      \begin{array}{cc}
        {\partial^2 S \over \partial {\bm \theta} \partial {\bm 
        \theta}^{\prime}}
        &  {\partial^2 S \over \partial {\bm \theta} \partial E} \\
        {\partial^2 S \over \partial E \partial {\bm \theta}^{\prime}}
        & {\partial^2 S \over \partial E^2} \\
      \end{array}
      \right|
\end{equation}
We assume that the Gaussian wave packet in Cartesian coordinates can be
define locally as a Gaussian in action-angle variables using the Wigner
function
\begin{equation}
      A_W({\bf I}, {\bm \theta}) = 2^d 
      \exp\{-(\Delta {\bf I})^2 / \hbar^2 - (\Delta {\bm \theta})^2 \}
\end{equation}
where $(\Delta {\bf I})_j \equiv \sigma_j ({\bf I} - {\bf I}_\alpha)_j$ and
$(\Delta {\bm \theta})_j \equiv ({\bm \theta} - {\bm \theta}_\alpha)_j
/ \sigma_j$. The inverse transform is evaluated by stationary phase to obtain
\begin{eqnarray}
      && \langle \alpha | {\bm \theta} \rangle \langle {\bm \theta}^\prime
        | \alpha \rangle = {1 \over \pi^{d/2} \prod \sigma_j} \\
        && \times \exp \left[ {-(\Delta {\bm \theta})^2 \over 2}
        - {(\Delta {\bm \theta'})^2 \over 2}
        + {i ({\bf I} - {\bf I}_\alpha)
        \cdot ({\bm \theta} - {\bm \theta}^\prime) \over \hbar} \right]
        \nonumber
\end{eqnarray}

The Gaussian localization in action-angle space implies that the dominant
contributions to the strength function arise from tori whose trajectories
are closed, i.e., rational tori ${\bf I_M}$.  An expansion about
the center of the Gaussian wave packet yields
\begin{equation}
      S_j({\bm \theta}, {\bm \theta}^\prime; E)
      = S_{\bf M}({\bm \theta}_\alpha, {\bm \theta}_\alpha; E)
      + {\bf I}_j \cdot ({\bm \theta} - {\bm \theta}^\prime)
\end{equation}
where quadratic terms have been neglected.  Using the above expansion of the
action, the integrals over the angles can be performed by stationary phase
giving
\begin{eqnarray}
      S_{\alpha, osc}(E, \lambda) &\approx& {2^{(d + 1)/2} \prod \sigma_j \over
        \pi^{1/2} \hbar^{(d + 1) / 2}}
        \sum_{\bf M} {}^\prime |D_s|^{1/2} \\
      &\times& \cos\left({S_{\bf M}\over \hbar}
        - {\eta_{\bf M} \pi \over 2} + {\beta_{\bf M} \pi \over 4}\right)
        \exp \left[ {-(\Delta {\bf I})^2 \over \hbar^2} \right] \nonumber
\end{eqnarray}

The above result for the oscillatory part of the strength function can be used
to obtain the variance of intensities as
\begin{eqnarray}
      \sigma^2_\alpha &=& {2 \pi \epsilon \over \overline{d}}
        \left \langle S^2_{\alpha, osc}(E, \lambda) \right \rangle_E 
        \nonumber \\
      &\approx& {\epsilon 2^{d + 1} \prod \sigma^2_j \over V \hbar}
        \left \langle \sum_{\bf M}{}^\prime |D_s|
        \exp \left[ {-2 (\Delta {\bf I})^2 \over \hbar^2} \right] \right. 
        \nonumber \\
      && \times \left. 
        \exp \left( {-2 \epsilon T_{\bf M} \over \hbar} \right)\right \rangle_E
\end{eqnarray}
where we have used the diagonal approximation for the sum over ${\bf M}$.
Since the trace of the Green function is the density of states given by
Berry and Tabor and the density of states is the energy derivative of the
spectral staircase function, then $|D_s| = T^2_{\bf M} A^2_{\bf M}
/ (2 \pi)^{d - 1}$.  Hence, given the Hannay-Ozorio sum rule for
$\sum_{\bf M} A^2_{\bf M}$, Eq.~(\ref{eq:sum_rule}), then
\begin{equation}
      \sum_{\bf M}{}^\prime |D_s| \cdots \rightarrow
      {V \over (2 \pi)^d} \int dT \cdots
\end{equation}
Since the Gaussian weighting of the action variables is independent of
the period, it follows that
\begin{equation}
      \sigma^2_\alpha \approx {\prod \sigma^2_j \over \pi^d}
      \left \langle \exp \left[ {-2 (\Delta {\bf I})^2
      \over \hbar^2}\right] \right \rangle_{\bf M}
\end{equation}

\section{Ballistic Vs. Diffusive Behavior}
\label{table}

We tried three different methods for separating out the ballistic
and diffusive orbits.  The fraction of ballistic orbits $f$ was deduced by
\begin{equation}
   f = {c_2(k) \over 4 \pi^2 \zeta(k)}
\end{equation}
where $c_2(k)$ was determined by Eq.~(\ref{eq:mixed}).
The first method was to divide the orbits by
a fixed value of the Lyapunov exponent.  It turned out that a Lyapunov
exponent of 0.5 gave fairly good result, but there is no explanation
of this value.  Table~\ref{table1} summarizes these results.
\begin{table}[!t]
   \caption{Fixed Lyapunov exponent cutoff. $f_1$ is the fraction of
   orbits that have a Lyapunov exponent less than $\mu = 0.5$.}
   \label{table1}
   \begin{ruledtabular}
     \begin{tabular}{|c|c|c|c|}
       $k$ & $\zeta(k)$ & $f$ & $f_1$ \\ \hline
       1.5 & $9.62 \times 10^{-5}$ & 0.982 & 1.000 \\
       2.0 & $1.11 \times 10^{-4}$ & 0.856 & 0.798 \\
       2.5 & $1.76 \times 10^{-4}$ & 0.358 & 0.336 \\
       3.0 & $3.66 \times 10^{-4}$ & 0.161 & 0.142 \\
       3.5 & $3.58 \times 10^{-4}$ & 0.166 & 0.146 \\
       4.0 & $3.67 \times 10^{-4}$ & 0.118 & 0.107 \\
     \end{tabular}
   \end{ruledtabular}
\end{table}
The next method was to take twice the fraction of stable orbits as the number
of ballistic orbits. This is summarized in Table~\ref{table2}.
\begin{table}[!t]
   \caption{Twice the regular orbits.  $f_2$ is twice the fraction of
   regular orbits and $\mu_2$ is the cutoff of the Lyapunov exponent
   associated with $f_2$.}
   \label{table2}
   \begin{ruledtabular}
     \begin{tabular}{|c|c|c|c|c|}
       $k$ & $\zeta(k)$ & $f$ & $f_2$ & $\mu_2$ \\ \hline
       1.5 & $1.17 \times 10^{-4}$ & 0.812 & 0.687 & 0.318 \\
       2.0 & $1.49 \times 10^{-4}$ & 0.638 & 0.526 & 0.443 \\
       2.5 & $2.19 \times 10^{-4}$ & 0.288 & 0.256 & 0.420 \\
       3.0 & $2.19 \times 10^{-4}$ & 0.269 & 0.241 & 0.590 \\
       3.5 & $2.21 \times 10^{-4}$ & 0.268 & 0.246 & 0.702 \\
       4.0 & $2.29 \times 10^{-4}$ & 0.189 & 0.179 & 0.713 \\
     \end{tabular}
   \end{ruledtabular}
\end{table}
For the final method we took the greatest number of orbits up to some
Lyapunov exponent that did not change $\zeta(k)$ significantly as the
number of ballistic orbits and is summarize is Table~\ref{table3}.
\begin{table}[!t]
   \caption{Stable quadratic.  $f_3$ is the number of orbits included
   in the ballistic term which does not change its coefficient significantly.
   $\mu_3$ is the corresponding Lyapunov exponent.}
   \label{table3}
   \begin{ruledtabular}
     \begin{tabular}{|c|c|c|c|c|}
       $k$ & $\zeta(k)$ & $f$ & $f_3$ & $\mu_3$ \\ \hline
       1.5 & $1.84 \times 10^{-4}$ & 0.516 & 0.412 & 0.2 \\
       2.0 & $2.58 \times 10^{-4}$ & 0.368 & 0.302 & 0.3 \\
       2.5 & $3.30 \times 10^{-4}$ & 0.191 & 0.165 & 0.2 \\
       3.0 & $4.26 \times 10^{-4}$ & 0.139 & 0.124 & 0.4 \\
       3.5 & $4.01 \times 10^{-4}$ & 0.148 & 0.137 & 0.4 \\
       4.0 & $4.31 \times 10^{-4}$ & 0.105 & 0.100 & 0.4 \\
     \end{tabular}
   \end{ruledtabular}
\end{table}


\begin{thebibliography}{}
\bibitem{note}  This scaling of the variance of the intensities differs
from our previous publication (Ref.~\cite{cerruti:01}), which contained
an error in the derivation of the average intensities and sum rule.
\end{thebibliography}
\end{document}